\documentclass[multphys,vecphys]{svmult}

\usepackage{makeidx}     
\usepackage{graphicx}    
\usepackage{multicol}    

\makeindex             

\begin{document}


\newcommand{\vnhat}{\hat{\vec{n}}}
\newcommand{\ve}{\vec{e}}
\newcommand{\vk}{\vec{k}}
\newcommand{\vv}{\vec{v}}
\newcommand{\vvb}{\vec{v}_{\mathrm{b}}}
\newcommand{\vvgam}{\vec{v}_\gamma}
\newcommand{\vp}{\vec{p}}
\newcommand{\vx}{\vec{x}}
\newcommand{\nel}{n_{\mathrm{e}}}
\newcommand{\sigT}{\sigma_{\mathrm{T}}}
\newcommand{\velb}{v_{\mathrm{b}}}
\newcommand{\velgam}{v_\gamma}
\newcommand{\dotvelb}{\dot{v}_{\mathrm{b}}}
\renewcommand{\Theta}{\varTheta}
\renewcommand{\Delta}{\varDelta}
\renewcommand{\Lambda}{\varLambda}
\renewcommand{\Omega}{\varOmega}



\title*{Anisotropies in the Cosmic Microwave Background}
\author{Anthony Challinor}
\institute{Astrophysics Group, Cavendish Laboratory, Madingley Road,\\
Cambridge, CB3 0HE,  U.K.
\texttt{a.d.challinor@mrao.cam.ac.uk}}
%
%
\maketitle

\begin{abstract}
The linear anisotropies in the temperature of the cosmic microwave
background (CMB) radiation and its polarization provide a clean picture of
fluctuations in the universe some $370\,\mathrm{kyr}$ after the big bang.
Simple physics connects these fluctuations with those present in the
ultra-high-energy universe, and this makes the CMB anisotropies a powerful
tool for constraining the fundamental physics that was responsible for the
generation of structure. Late-time effects also leave their mark, making the
CMB temperature and polarization useful probes of dark energy
and the astrophysics of reionization. In this review we discuss the simple
physics that processes primordial perturbations into the linear temperature
and polarization anisotropies. We also describe the role of the CMB in
constraining cosmological parameters, and review some of the highlights of
the science extracted from recent observations and the implications of this
for fundamental physics.
\end{abstract}

\section{Introduction}
\label{adc:sec:intro}

The cosmic microwave background (CMB) radiation has played an essential role
in shaping our current understanding of the large-scale properties of the
universe. The discovery of this radiation in 1965 by Penzias and
Wilson~\cite{adc:pw65}, and its subsequent interpretation as the relic
radiation from a hot, dense phase of the universe~\cite{adc:dicke65}
put the hot big bang model on
a firm observational footing. The prediction of angular variations in the
temperature of the radiation, due to the propagation of photons through an
inhomogeneous universe, followed shortly after~\cite{adc:sw67}, but it was not
until 1992 that these were finally detected by the Differential Microwave
Radiometers (DMR) experiment on the Cosmic Background Explorer (COBE)
satellite~\cite{adc:smoot92}. The fractional temperature anisotropies are
at the level of
$10^{-5}$, consistent with structure formation in cold dark matter (CDM)
models~\cite{adc:peebles82,adc:bond84}, but much smaller than earlier
predictions for baryon-dominated universes~\cite{adc:sw67,adc:peebles70}.
Another experiment on COBE, the Far InfraRed Absolute
Spectrophotometer (FIRAS), spectacularly confirmed the black-body spectrum
of the CMB and determined the (isotropic) temperature to be
2.725\,K~\cite{adc:mather94,adc:mather99}.

In the period since COBE, many experiments have mapped the CMB anisotropies
on a range of angular scales from degrees to arcminutes (see \cite{adc:bond03}
for a recent review), culminating in the first-year release of all-sky
data from the Wilkinson Microwave Anisotropy Probe (WMAP) satellite in February
2003~\cite{adc:bennett03}. The observed modulation in the amplitude of 
the anisotropies with angular scale is fully consistent with predictions
based on coherent, acoustic oscillations~\cite{adc:peebles70}, derived from
gravitational
instability of initially adiabatic density perturbations in a universe with
nearly-flat spatial sections. The amplitude and scale of these acoustic
features has allowed many of the key cosmological
parameters to be determined with unprecedented precision~\cite{adc:spergel03},
and a strong concordance with other cosmological probes has emerged.

In this review we describe the essential physics of the temperature
anisotropies of the CMB, and its recently-detected
polarization~\cite{adc:kovac02},
and discuss how these are used to constrain cosmological models.
For reviews that are similar in spirit, but from the pre-WMAP era
see e.g.~\cite{adc:hu02,adc:hu03}.
We begin in Sect.~\ref{adc:sec:fundamentals} with
the fundamentals of CMB physics, presenting the kinetic theory of the
CMB in an inhomogeneous universe, and the various physical mechanisms
that process initial fluctuations in the distribution of matter and
spacetime geometry into temperature anisotropies. Section~\ref{adc:sec:params}
discusses the effect of cosmological parameters on the power
spectrum of the temperature anisotropies, and the limits to parameter
determination from the CMB alone. The physics of CMB polarization is
reviewed in Sect.~\ref{adc:sec:pol}, and the additional information that
polarization brings over temperature anisotropies alone is considered.
Finally, in Sect.~\ref{adc:sec:highlights} we describe some of the scientific
highlights that have emerged
from recent CMB observations, including the detection of CMB polarization,
implications for inflation, and the direct signature of dark-energy through
correlations between the large-scale anisotropies and tracers of the mass
distribution in the local universe.
Throughout, we illustrate our discussion with
computations based on $\Lambda$CDM cosmologies, with baryon density
$\Omega_\mathrm{b}h^2 = 0.023$ and
cold dark matter density $\Omega_\mathrm{c} h^2 = 0.111$.
For flat models we take the dark-energy density parameter to be
$\Omega_\Lambda = 0.75$ giving a Hubble parameter $H_0 = 73\,
\mathrm{km\,s}^{-1}\,\mathrm{Mpc}^{-1}$. We adopt units with $c=1$ throughout,
and use a spacetime metric signature $+---$.

\section{Fundamentals of CMB Physics}
\label{adc:sec:fundamentals}

In this section we aim to give a reasonably self-contained review of the
essential elements of CMB physics. 

\subsection{Thermal History and Recombination}
\label{adc:subsec:thermal}

The high temperature of the early universe maintained a low
equilibrium fraction of neutral atoms, and a correspondingly high
number density of free electrons. Coulomb scattering between the ions
and electrons kept them in local kinetic equilibrium, and Thomson scattering
of photons tended to maintain the isotropy of the CMB in the baryon rest frame.
As the universe expanded and cooled, the dominant element hydrogen started to
recombine when the temperature fell below $\sim$ 4000\,K -- a factor of
40 lower than might be anticipated from the
13.6-eV inoization potential of hydrogen,
due to the large ratio of the number of photons to baryons. The details
of recombination are complicated since the processes that give rise to
net recombination occur too slowly to maintain chemical equilibrium between
the electrons, protons and atoms during the later stages of
recombination~\cite{adc:peebles68,adc:zeldovich69} (see~\cite{adc:seager00}
for recent refinements). The most
important quantity for CMB anisotropy formation is the visibility function --
the probability that a photon last scattered as a function of time.
The visibility function peaks around $\sim 370\,\mathrm{kyr}$ after the
big bang, and has a width $\sim 115\,\mathrm{kyr}$, a small
fraction of the current age $\sim 13.5 \,\mathrm{Gyr}$~\cite{adc:spergel03}.
After recombination, photons travelled mostly unimpeded through the
inhomogeneous universe, imprinting fluctuations in the radiation temperature,
the gravitational potentials, and the bulk velocity of the radiation where they
last scattered, as the temperature anisotropies that we observe today.
A small fraction of CMB photons (current results from CMB polarization
measurements~\cite{adc:kogut03} indicate around 20 per cent;
see also Sect.~\ref{adc:subsec:polhighlights}) underwent further
scattering once the universe reionized due to to the ionizing flux from
the first non-linear structures.

\subsection{Statistics of CMB Anisotropies}
\label{adc:subsec:statistics}

The spectrum of the CMB brightness along any direction $\vnhat$ is very nearly
thermal with a temperature $T(\vnhat)$. The temperature depends only weakly
on direction, with fluctuations $\Delta T(\vnhat)$ at the level of
$10^{-5}$ of the average temperature $T=2.725$\,K. It is convenient to
expand the temperature fluctuation in spherical harmonics,
\begin{equation}
\Delta T(\vnhat) / T = \sum_{lm} a_{lm} Y_{lm}(\vnhat) \; ,
\label{adc:eq1}
\end{equation}
with $a_{lm}^* = (-1)^m a_{l-m}$ since the temperature is a real field.
The sum in (\ref{adc:eq1}) runs over $l \geq 1$, but the dipole ($l=1$) is
usually removed explicitly when analysing data since it depends linearly on
the velocity of the observer. Multipoles at $l$ encode spatial information
with characteristic angular scale $\sim \pi/l$.

\begin{figure}[t!]
\centering
\includegraphics[height=10cm,angle=90]{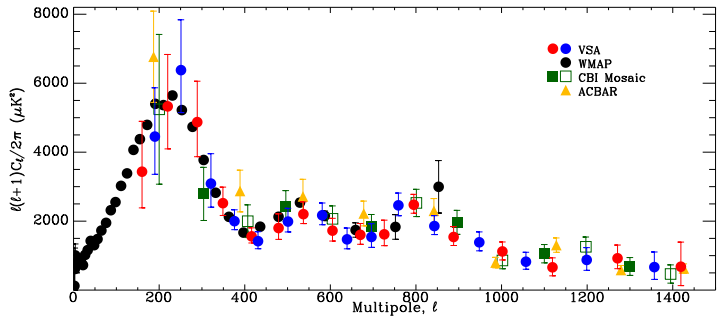}
\caption{Compilation of CMB anisotropy measurements (as of February 2004)
from WMAP (black filled circles),
the Very Small Array (VSA~\cite{adc:dickinson04};
shaded circles representing two interleaving
binning schemes), the Cosmic Background Imager
(CBI~\cite{adc:mason03,adc:pearson03}; open and filled squares
for two different binning schemes) and the Arcminute Cosmology Bolometer
Array Receiver (ACBAR~\cite{adc:kuo04}; triangles).
(Figure reproduced, with permission, from~\cite{adc:dickinson04}.)}
\label{adc:fig1} 
\end{figure}

The statistical properties of the fluctuations in a perturbed cosmology
can be expected to respect the symmetries of the background
model. In the case of Robertson--Walker models,
the rotational symmetry of the
background ensures that the multipoles $a_{lm}$ are uncorrelated for different
values of $l$ and $m$:
\begin{equation}
\langle a_{lm} a^*_{l'm'} \rangle = C_l \delta_{ll'} \delta_{mm'} \; ,
\label{adc:eq2}
\end{equation}
which defines the power spectrum $C_l$. The angle brackets in this equation
denote the average over an ensemble of realisations of the fluctuations.
The simplest models of inflation predict that the fluctuations should also be
Gaussian at early times, and this is preserved by linear evolution of the
small fluctuations. If Gaussian, the $a_{lm}$s are also independent, and
the power spectrum provides the complete statistical description of the
temperature anisotropies. For this reason, measuring the anisotropy
power spectrum has, so far, been the main goal of observational CMB research.
Temperature anisotropies have now been detected up to $l$ of a few thousand;
a recent compilation of current data as of February 2004 is given in
Fig.~\ref{adc:fig1}.

The correlation between the temperature anisotropies along two directions
evaluates to
\begin{equation}
\langle \Delta T(\vnhat_1) \Delta T(\vnhat_2) \rangle
= T^2 \sum_l \frac{2l+1}{4\pi} C_l P_l(\cos\theta)\;,
\label{adc:eq3}
\end{equation}
which depends only on the angular separation $\theta$ as required by rotational
invariance. Here, $P_l(x)$ are the Legendre polynomials. The mean-square
temperature anisotropy is
\begin{equation}
\langle \Delta T ^2 \rangle
= T^2 \sum_l \frac{2l+1}{4\pi} C_l
\approx T^2 \int \frac{l(l+1)}{2\pi} C_l \, \D \ln l\; ,
\label{adc:eq4}
\end{equation}
so that the quantity $l(l+1)C_l /2\pi$, which is conventionally plotted,
is approximately the power per decade in $l$ of the temperature anisotropies.

\subsection{Kinetic Theory}
\label{adc:subsec:kinetic}

The CMB photons can be described by a one-particle distribution function
$f(x^a,p^a)$ that is a function of the spacetime position $x^a$ and
four-momentum $p^a$ of the photon. It is defined such that the number of
photons contained in a proper three-volume element $\D^3 \vx$ and with
three-momentum in $\D^3 \vp$ is $f \D^3 \vx \D^3 \vp$. The phase-space volume
element $\D^3 \vx \D^3 \vp$ is Lorentz-invariant and is conserved along the
photon path through phase space (see, e.g.~\cite{adc:mtw}).
It follows that $f$ is also frame-invariant, and is conserved in the absence
of scattering. To calculate the anisotropies in the CMB temperature, we must
evolve the photon distribution function in the perturbed universe.

To avoid over-complicating our discussion, we shall only consider
spatially-flat models here, and, for the moment, ignore the effects of
polarization. For a more complete discussion, including these complications,
see e.g.~\cite{adc:hu98,adc:challinor00}. Curvature mostly affects
the CMB through the geometrical projection of linear scales at last scattering
to angular scales on the sky today, but has a negligible impact on
pre-recombination physics and hence much of the discussion in this section.
The subject of cosmological
perturbation theory is rich in methodology, but, for pedagogical reasons, we
adopt here the most straightforward approach which is to work directly with
the metric perturbations. This is also the most prevalent in the CMB
literature. The 1+3-covariant approach~\cite{adc:ellis89} is a well-developed
alternative that is arguably more physically-transparent than metric-based
techniques. It has also been applied extensively in the context of CMB
physics~\cite{adc:challinor00,adc:challinor99,adc:challinor00b,adc:maartens99,adc:gebbie00a,adc:gebbie00b}. The majority of our discussion will be of scalar
perturbations, where all perturbed three-tensors can be derived from the
spatial derivatives of scalar functions, although we discuss tensor
perturbations briefly in Sect.~\ref{adc:subsec:tensors}.

For scalar
perturbations in spatially-flat models we can choose a gauge such that the
spacetime metric is~\cite{adc:ma95}
\begin{equation}
\D s^2 = a^2(\eta)[(1+2\psi)\D\eta^2 - (1-2\phi)\D\vx^2]\; ,
\label{adc:eq5}
\end{equation}
where $\eta$ is conformal time (related to proper time
$t$ by $\D t = a \D\eta$), $a$ is the scale factor in the background model
and, now, $\vx$ is comoving position. This gauge, known as the
conformal Newtonian or longitudinal gauge, has the property that
the congruence of worldlines with constant $\vx$ have zero shear. The two
scalar potentials $\phi$ and $\psi$ constitute the scalar perturbation to the
metric, with $\phi$ playing a similar role to the Newtonian gravitational
potential. In the absence of anisotropic stress, $\phi$ and $\psi$ are equal.
We parameterise the photon four-momentum with its energy $\epsilon/a$
and direction $\ve$ (with $\ve^2 = 1$), as seen by an observer at constant
$\vx$, so that
\begin{equation}
p^\mu = a^{-2} \epsilon [1-\psi,(1+\phi)\ve]\; .
\label{adc:eq6}
\end{equation}
Free photons move on the geodesics of the perturbed metric,
$p^\mu \nabla_\mu p^\nu = 0$, so the energy and direction evolve as
\begin{eqnarray}
\D \epsilon / \D\eta &=& - \epsilon \D\psi / \D\eta + \epsilon (\dot{\phi}
+ \dot{\psi})\; , \label{adc:eq7} \\ 
\D \ve / \D\eta &=& - \vec{\nabla}_{\perp} (\phi + \psi)\; ,
\label{adc:eq8}
\end{eqnarray}
where dots denote $\partial / \partial \eta$ and $\vec{\nabla}_{\perp}$ is the
three-gradient projected perpendicular to $\ve$. We see immediately that
$\epsilon$ is conserved in the absence of perturbations, so that the energy
redshifts in proportion to the scale factor in the background model. The
change in direction of the photon due to the projected gradient of the
potentials in the perturbed universe gives rise to gravitational
lensing (see e.g.~\cite{adc:bartelmann01} for a review).

The dominant scattering mechanism to affect CMB anisotropies is classical
Thomson scattering off free electrons, since around recombination the average
photon energy is small compared to the rest mass of the electron. Furthermore,
the thermal distribution of electron velocities can be ignored due to the low
temperature. The evolution of the photon distribution function in the
presence of Thomson scattering is
\begin{eqnarray}
\frac{\D f}{\D \eta} &=& - a(1+\psi) \nel \sigT f + \frac{3}{16\pi} a(1+\psi)
\nel\sigT \int f(\epsilon,\ve')[(1+(\ve\cdot\ve')^2]\, \D\ve' \nonumber \\
&&\mbox{} - a \nel\sigT
\ve \cdot \vvb \epsilon \frac{\partial f}{\partial \epsilon}\; ,
\label{adc:eq9}
\end{eqnarray}
where $\nel$ is the electron (proper) number density, $\sigT$ is the Thomson
cross section, and the electron peculiar velocity is $\vvb = \D \vx / \D \eta$.
The derivative on the left of (\ref{adc:eq9}) is along the photon path in
phase space:
\begin{equation}
\frac{\D f}{\D \eta} = \frac{\partial f}{\partial \eta} + \ve \cdot
\vec{\nabla} f + (\dot{\phi} - \ve \cdot \vec{\nabla} \psi) \epsilon
\frac{\partial f}{\partial \epsilon}
\label{adc:eq10}
\end{equation}
to first order, where we have used (\ref{adc:eq7}) and (\ref{adc:eq8}) and
the fact that the anisotropies of $f$ are first order.
The first term on the right of (\ref{adc:eq9}) describes scattering out of the
beam, and the second scattering into the beam. The final term arises from
the out-scattering of the additional dipole moment in the distribution function
seen by the electrons due to the Doppler effect. In the background model
$f$ is isotropic and the net scattering term vanishes, so that $f$ is a
function of the conserved $\epsilon$ only: $f=\bar{f}(\epsilon)$. Thermal
equilibrium ensures that $\bar{f}$ is a Planck function. 

The fluctuations in the photon distribution function inherit an energy
dependence $\epsilon \partial \bar{f}/\partial \epsilon$ from the source
terms in the Boltzmann equation~(\ref{adc:eq9}). Separating out the
background contribution to $f$, and its energy dependence, we can write
\begin{equation}
f(\eta,\vx,\epsilon,\ve) = \bar{f}(\epsilon)[1-\Theta(\eta,\vx,\ve)\D \ln
\bar{f} / \D \ln \epsilon],
\label{adc:eq11}
\end{equation}
so that the CMB spectrum is Planckian but with a direction-dependent
temperature $\Delta T / T = \Theta$. Using the Lorentz invariance of $f$, it
is not difficult to show that the quadrupole and higher moments of $\Theta$
are gauge-invariant.
If we now substitute for $f$ in (\ref{adc:eq9}), we find the Boltzmann equation
for $\Theta$:
\begin{eqnarray}
\frac{\partial (\Theta+\psi)}{\partial \eta} + \ve \cdot \vec{\nabla}
(\Theta + \psi) &=& - a \nel \sigT \Theta + \frac{3}{16\pi}
a \nel\sigT \int \Theta(\ve')[(1+(\ve\cdot\ve')^2]\, \D\ve' \nonumber \\
&&\mbox{} + a \nel\sigT \ve \cdot \vvb + \dot{\phi} + \dot{\psi} \; .
\label{adc:eq12}
\end{eqnarray}
The formal solution of this equation is an integral along the line of sight
$\vnhat = -\ve$,
\begin{equation}
[\Theta (\vnhat) + \psi]_R = \E^{-\tau} [\Theta (\vnhat) + \psi)]_E +
\int_E^R \E^{-\tau} S\, \D\eta \; ,
\label{adc:eq13}
\end{equation}
where $R$ is the reception event, $E$ is the emission event, and
$\tau \equiv \int a\nel\sigT \, \D\eta$ is the optical depth back from $R$.
The source term $S$ is given by the right-hand side of (\ref{adc:eq12}), but
with $\Theta$ replaced by $-\psi$ in the first term.

We gain useful insight into the physics of anisotropy formation by
approximating the last scattering surface as sharp (which is harmless on
large angular scales), and ignoring the quadrupole CMB anisotropy at last
scattering. In this case (\ref{adc:eq13}) reduces to
\begin{equation}
[\Theta (\vnhat) + \psi]_R = \Theta_0 |_E + \psi |_E
- \vnhat \cdot \vvb |_E + \int_E^R (\dot{\psi}+\dot{\phi})\, \D\eta\; , 
\label{adc:eq14}
\end{equation}
where $\Theta_0$ is the isotropic part of $\Theta$, and is proportional to the
fluctuation in the photon energy density.
The various terms in this equation have a simple physical interpretation.
The temperature received along direction $\vnhat$ is the isotropic
temperature of the CMB at the last scattering event on the line of sight,
$\Theta_0$, corrected for the gravitational redshift due to the
difference in potential between $E$ and $R$, and the Doppler shift
$\ve \cdot \vvb|_E$ resulting from scattering off moving electrons. Finally,
there is an additional gravitational redshift contribution arising from
evolution of the gravitational potentials~\cite{adc:sw67}.

\subsubsection{Machinery for an Accurate Calculation}
\label{adc:subsubsec:machine}

An accurate calculation of the CMB anisotropy on all scales where linear
perturbation theory is valid requires a full numerical solution of the
Boltzmann equation. The starting point is to expand $\Theta(\theta,\vx,\ve)$
in appropriate basis functions. For scalar perturbations, these are the
contraction of the (irreducible) trace-free tensor products $e^{\langle i_1}
\dots e^{i_l \rangle}$ (the angle brackets denoting the trace-free part)
with trace-free (spatial) tensors derived from
derivatives of scalars~\cite{adc:challinor99,adc:gebbie00a,adc:wilson83}.
Fourier expanding the scalar functions,
we end up forming contractions between $e^{\langle i_1} \dots e^{i_l \rangle}$
and $\hat{k}_{\langle i_1} \dots \hat{k}_{i_l \rangle}$ where $\hat{\vk}$ is
the wavevector. These contractions reduce to Legendre polynomials of
$\hat{\vk}\cdot \ve$, and so the normal-mode expansion of $\Theta$ for
scalar perturbations takes the form
\begin{equation}
\Theta(\eta,\vx,\ve) = \sum_{l\geq 0} \int \frac{\D^3 \vk}{(2\pi)^{3/2}}
(-\I)^l \Theta_l(\eta,\vk) P_l(\hat{\vk}\cdot\ve) \E^{\I \vk \cdot \vx}\; . 
\label{adc:eq15}
\end{equation}
It is straightforward to show that the implied azimuthal symmetry about the
wavevector is consistent with the Boltzmann equation~(\ref{adc:eq12}).
Inserting the expansion of $\Theta$ into this equation gives the Boltzmann
hierarchy for the moments $\Theta_l$:
\begin{eqnarray}
\dot{\Theta}_l + k \left(\frac{l+1}{2l+3}\Theta_{l+1} - \frac{l}{2l-1}
\Theta_{l-1}\right) &=& a\nel\sigT\left[(\delta_{l0}-1)\Theta_l
- \delta_{l1}\velb + \frac{1}{10} \Theta_2\right] \nonumber \\
&&\mbox{} + \delta_{l0} \dot{\phi} + \delta_{l1} k\psi \; ,
\label{adc:eq16}
\end{eqnarray}
where $\vvb = \int \I \hat{\vk} \velb(\vk) \E^{\I\vk\cdot\vx}
\,\D^3\vk/(2\pi)^{3/2}$, and $\phi$ and $\psi$ are the Fourier transforms
of the potentials.
This system of ordinary differential equations
can be integrated directly with the linearised Einstein equations for the
metric perturbations, and the fluid equations governing perturbations in the
other matter components, as in the publically-available COSMICS
code~\cite{adc:ma95}. Careful treatment of the truncation of the hierarchy
is necessary to avoid unphysical reflection of power back down through the
moments.

A faster way to solve the Boltzmann equation numerically is to use the
line-of-sight solution (\ref{adc:eq13}), as in the widely-used CMBFAST
code~\cite{adc:seljak96} and its parallelised derivative
CAMB~\cite{adc:lewis00}. Inserting the expansion~(\ref{adc:eq15}) gives
the integral solution to the hierarchy
\begin{eqnarray}
\Theta_l|_{\eta_0} &=& (2l+1) \int_0^{\eta_0}\D\eta\, \E^{-\tau} \Big[
(\dot{\phi}+\dot{\psi})j_l(k\Delta \eta) - \dot{\tau}(\Theta_0 + \psi)
j_l(k\Delta \eta) \nonumber \\
&&\phantom{(2l+1) \int_0^{\eta_0}\D\eta\, \E^{-\tau}}
+ \dot{\tau} \velb j_l'(k\Delta \eta) - \frac{1}{20}
\dot{\tau}\Theta_2(3j_l''+ j_l)(k\Delta \eta)\Big]\; , 
\label{adc:eq17}
\end{eqnarray}
where $\Delta \eta \equiv \eta_0 - \eta$, $j_l$ is a spherical Bessel function,
and primes denote derivatives with respect to the argument. Using the
integral solution, it is only necessary to evolve the Boltzmann hierarchy
to modest $l$ to compute accurately the source terms that appear in the
integrand. The integral approach is thus significantly faster than a direct
solution of the hierarchy.

The spherical multipoles $a_{lm}$ of the temperature anisotropy can be
extracted from (\ref{adc:eq15}) as
\begin{equation}
a_{lm} = 4\pi \I^l \int \frac{\D^3 \vk}{(2\pi)^{3/2}}\, \frac{\Theta_l}{2l+1}
Y_{lm}^*(\hat{\vk}) \E^{\I\vk\cdot \vx}\; .
\label{adc:eq18}
\end{equation}
Statistical homogeneity and isotropy imply that the equal-time correlator
\begin{equation}
\langle \Theta_l(\eta,\vk) \Theta_l^*(\eta,\vk') \rangle = \frac{2\pi^2}{k^3}
\Theta_l^2(\eta,k) \delta(\vk-\vk')\; ,
\label{adc:eq19}
\end{equation}
so forming the correlation $\langle a_{lm} a^*_{l'm'} \rangle$ gives the
power spectrum
\begin{equation}
C_l = \frac{4\pi}{(2l+1)^2} \int \Theta_l^2(k) \, \D\ln k\; .
\label{adc:eq20}
\end{equation}
If we consider (pure) perturbation modes characterised by a single independent
stochastic amplitude per Fourier mode (such as the comoving curvature for
the adiabatic
mode; see Sect.~\ref{adc:subsec:dynamics}), the power $\Theta_l^2(k)$ is
proportional to the power spectrum of that amplitude. The spherical Bessel
functions in (\ref{adc:eq17}) peak sharply at $k\Delta \eta = l$ for large $l$,
so that multipoles $l$ are mainly probing spatial structure with wavenumber
$k \sim l / \Delta \eta$ at last scattering. The oscillatory tails of the
Bessel functions mean that some power from a given $k$ does also enter larger
scale anisotropies. Physically, this arises from Fourier modes that are not
aligned with their wavevector perpendicular to the line of sight. As we discuss
in the next section, the tightly-coupled system of photons and baryons
undergoes acoustic oscillations prior to recombination on scales inside the
sound horizon. For the pure perturbation modes, all modes with a given
wavenumber reach the maxima or minima of their oscillation at the same time,
irrespective of the direction of $\vk$, and so we expect modulation in the
$C_l$s on sub-degree scales. The first three of these acoustic peaks have now
been measured definitively; see Fig.~\ref{adc:fig1}.

\subsection{Photon--Baryon Dynamics}
\label{adc:subsec:dynamics}

Prior to recombination, the mean free path of CMB photons is
$\sim 4.9\times 10^4 (\Omega_\mathrm{b} h^2)^{-1}(1+z)^{-2}$\,Mpc. On comoving
scales below this length the photons and baryons behave as a tightly-coupled
fluid, with the CMB almost isotropic in the baryon frame. In this limit, only
the $l=0$ and $l=1$ moments of the distribution function are significant.

The stress-energy tensor of the photons is given in terms of the distribution
function by
\begin{equation}
T^{\mu\nu} = a^{-2} \int f(\eta,\vx,\epsilon,\ve) p^\mu p^\nu \epsilon \,
\D\epsilon \D\ve\; ,
\label{adc:eq21}
\end{equation}
so that the Fourier modes of the fractional over-density of the photons
are $\delta_\gamma = 4\Theta_0$ and the photon (bulk) velocity
$v_\gamma = -\Theta_1$. The anisotropic stress is proportional to $\Theta_2$.
In terms of these variables, the first two moment equations of the Boltzmann
hierarchy become
\begin{eqnarray}
\dot{\delta}_\gamma - \frac{4}{3} k v_\gamma - 4\dot{\phi} & = & 0\; ,
\label{adc:eq22}\\
\dot{v}_\gamma + \frac{1}{4} k \delta_\gamma - \frac{2}{5} k\Theta_2 + k\psi
&=& \dot{\tau}(v_\gamma - \velb)\; . \label{adc:eq23}
\end{eqnarray}
Here, the derivative of the optical depth $\dot{\tau} = - a\nel\sigT$ (and so
is negative).
The momentum exchange between the photons and baryons due to the
drag term in (\ref{adc:eq23}) gives rise to a similar term in the Euler
equation for the baryons:
\begin{equation}
\dotvelb + \mathcal{H} \velb + k\psi = R^{-1} \dot{\tau} (\velb - v_\gamma)\; ,
\label{adc:eq24}
\end{equation}
where we have ignored baryon pressure. The ratio of the baryon energy density
to the photon enthalpy is $R\equiv 3 \rho_\mathrm{b}/
4 \rho_\gamma$ and is proportional to the scale factor $a$, and $\mathcal{H}
\equiv \dot{a}/a$ is the conformal Hubble parameter.

In the tightly-coupled limit $|\dot{\tau}^{-1}| \ll k^{-1}$ and
$\mathcal{H}^{-1}$. In this limit, we can treat the ratios of the mean-free
path to the wavelength and the Hubble time as small perturbative parameters.
Equations (\ref{adc:eq23}) and (\ref{adc:eq24}) then imply
that $v_\gamma = \velb$ to first order in the small quantities
$k / |\dot{\tau}|$ and $\mathcal{H} / |\dot{\tau}|$. Comparing the continuity
equation for the baryons,
\begin{equation}
\dot{\delta}_\mathrm{b} - k\velb - 3 \dot{\phi} = 0\; ,
\label{adc:eq25}
\end{equation}
with that for the photons, we see that
$\dot{\delta}_\gamma = 4 \dot{\delta}_\mathrm{b}/3$, so the \emph{evolution}
of the photon--baryon fluid is adiabatic, preserving the local ratio of the
number densities of photons to baryons. Combining (\ref{adc:eq23}) and
(\ref{adc:eq24}) to eliminate the scattering terms, and then using
$v_\gamma = \velb$, we find the evolution of the photon velocity to leading
order in tight coupling:
\begin{equation}
\dot{v}_\gamma + \frac{R}{1+R} \mathcal{H} v_\gamma + \frac{1}{4(1+R)}
k \delta_\gamma + k\psi = 0\; .
\label{adc:eq26}
\end{equation}
The $l > 1$ moments of the photon distribution function arise from the
balance between isotropisation by scattering and their generation by
photons free streaming over a mean free path; these moments are
suppressed by factors $(k/ |\dot{\tau}|)^{l-1}$. In particular, during
tight coupling $\Theta_2 \approx (20/27) k\dot{\tau}^{-1} v_\gamma$ ignoring
polarization. (The factor $20/27$ rises to $8/9$ if we correct for
polarization~\cite{adc:kaiser83}.)

Combining (\ref{adc:eq26}) with the photon continuity
equation (\ref{adc:eq22}) shows that the tightly-coupled dynamics of
$\delta_\gamma$ is that of a damped, simple-harmonic oscillator driven by
gravity~\cite{adc:hu95a}:
\begin{equation}
\ddot{\delta}_\gamma + \frac{\mathcal{H}R}{1+R} \dot{\delta}_\gamma
+ \frac{1}{3(1+R)}k^2 \delta_\gamma  = 4 \ddot{\phi} +
\frac{4 \mathcal{H}R}{1+R} \dot{\phi} - \frac{4}{3} k^2 \psi\; .
\label{adc:eq27}
\end{equation}
The damping term arises from the redshifting of the baryon momentum in
an expanding universe, while photon pressure provides the restoring force which
is weakly suppressed by the additional inertia of the baryons. The WKB
solutions to the homogeneous equation are
\begin{equation}
\delta_\gamma = (1+R)^{-1/4} \cos k r_\mathrm{s}\; , \phantom{xx}
\mathrm{and} \phantom{xx} \delta_\gamma = (1+R)^{-1/4} \cos k r_\mathrm{s}\; ,
\label{ac:eq28}
\end{equation}
where the \emph{sound horizon} $r_\mathrm{s} \equiv \int_0^\eta \D \eta' /
\sqrt{3(1+R)}$. Note also that for static potentials, and ignoring the
variation of $R$ with time, the mid-point of the
oscillation of $\delta_\gamma$ is shifted to $-4 (1+R)\psi$. The dependence of
this shift on the baryon density produces a baryon-dependent modulation of the
height of the acoustic peak in the temperature anisotropy power spectrum; see
Section~\ref{adc:sec:params}.

\begin{figure}[t!]
\centering
\includegraphics[height=8cm,angle=-90]{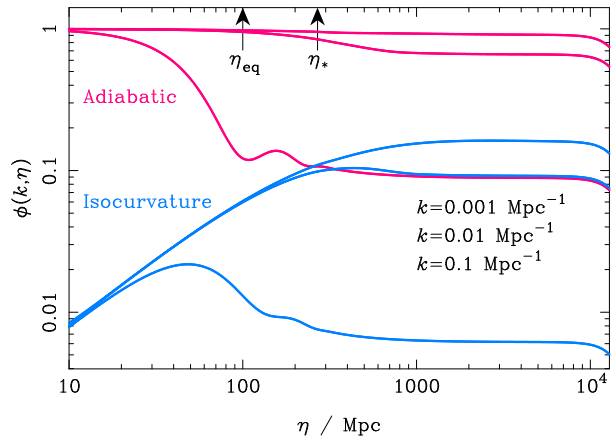}
\caption{Evolution of the potential $\phi$ in adiabatic and CDM-isocurvature
models for wavenumbers $k=0.001$, $0.01$ and $0.1\, \mathrm{Mpc}^{-1}$
(top to bottom respectively in matter domination). The conformal time at
matter--radiation equality $\eta_\mathrm{eq}$ and last scattering $\eta_*$
are marked by arrows.}
\label{adc:fig2} 
\end{figure}

The driving term in (\ref{adc:eq27}) depends on the evolution of the
gravitational potentials. If we ignore anisotropic stress, $\phi$ and $\psi$
are equal, and their Fourier modes evolve as
\begin{eqnarray}
\ddot{\phi} + 3 \mathcal{H}\left(1 + \frac{\dot{p}}{\dot{\rho}}\right)
\dot{\phi} + \left[2\dot{\mathcal{H}} + \left(1+3\frac{\dot{p}}
{\dot{\rho}}\right) \mathcal{H}^2\right]\phi &+&
\frac{\dot{p}}{\dot{\rho}} k^2 \phi \nonumber \\
&=& \frac{1}{2}\kappa a^2 \left(\delta p - \frac{\dot{p}}{\dot{\rho}}
\delta \rho \right)
\label{adc:eq29}
\end{eqnarray}
in a flat universe,
which follows from the perturbed Einstein field equations. Here, $\rho$ and
$p$ are the total density and pressure in the background model, $\delta
\rho$ and $\delta p$ are the Fourier modes of their perturbations, and
$\kappa \equiv 8\pi G$. The source
term is gauge-invariant; it vanishes for mixtures of barotropic fluids
[$p_i=p_i(\rho_i)$] with $\delta \rho_i / (\rho_i + p_i)$ the same for all
components. For \emph{adiabatic} perturbations, this latter condition holds
initially and is preserved on super-Hubble scales. It is also preserved in
the tightly-coupled photon--baryon fluid as we saw above. For adiabatic
perturbations, the potential is constant on scales larger than the sound
horizon when $p/\rho$ is constant, but decays during transitions in the
equation of state, such as from matter to radiation domination.
Above the sound horizon in flat models, it can be shown that the
quantity
\begin{equation}
\mathcal{R} \equiv - \phi - 2\frac{\mathcal H \dot{\phi} + \mathcal{H}^2}{%
\kappa a^2 (\rho + p)}
\label{adc:eq30}
\end{equation}
is conserved even through such transitions. The perturbation to the intrinsic
curvature of comoving hypersurfaces (i.e.\ those perpendicular to the the
four-velocity of observers who see no momentum density) is given in terms
of $\mathcal{R}$ as $4 (k^2 / a^2) \mathcal{R}$.
Using the constancy of $\mathcal{R}$ on large
scales, the potential falls by a factor of $9/10$ during the transition from
radiation to matter domination.
The evolution of the potential is illustrated in Fig.~\ref{adc:fig2} in a
flat $\Lambda$CDM model with parameters given in Sect.~\ref{adc:sec:intro}.
The potential oscillates inside
the sound horizon during radiation domination since the photons, which are
the dominant component at that time, undergo acoustic oscillations on such
scales. 

The behaviour of the potentials for \emph{isocurvature} perturbations is
quite different on large scales during radiation domination~\cite{adc:hu95b},
since the source term in (\ref{adc:eq29}) is then significant.
In isocurvature fluctuations, the initial perturbations in the energy densities
of the various components compensate each other in such a way that the
comoving curvature $\mathcal{R}=0$. Figure~\ref{adc:fig2} shows the
evolution of CDM-isocurvature modes, in which there is initially
a large fractional perturbation in the dark matter density, with a small
compensating fractional perturbation in the radiation. (The full set of
possibilities for regular isocurvature modes are discussed
in~\cite{adc:bucher00}.)
On large scales in radiation domination the potential grows as $a$, the
scale factor.

\subsubsection{Adiabatic Fluctuations}

\begin{figure}[t!]
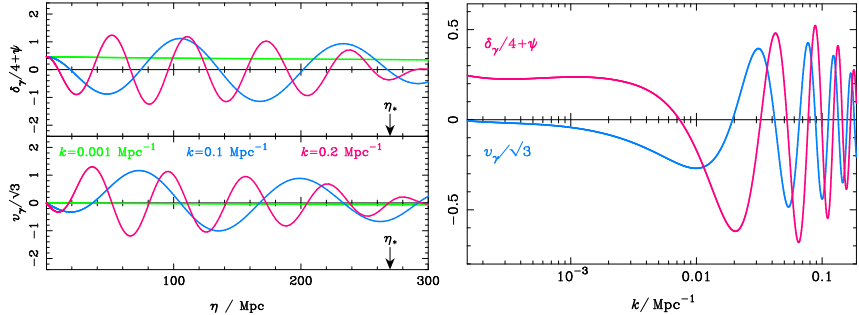

\begin{center}
\includegraphics[width=0.35\textwidth,angle=-90]{adc_fig3a.ps}
\includegraphics[width=0.35\textwidth,angle=-90]{adc_fig3b.ps}
\end{center}
\caption{Evolution of the combination $\delta_\gamma / 4 + \psi$ (top left)
and the photon velocity $v_\gamma$ (bottom left) which determine the
temperature anisotropies produced at last scattering (denoted by the arrow
at $\eta_*$). Three modes are
shown with wavenumbers $k=0.001$, $0.1$ and $0.2\,\mathrm{Mpc}^{-1}$, and the
initial conditions are adiabatic. The fluctuations at the time of last
scattering are shown as a function of linear scale in the right-hand plot.
}
\label{adc:fig3} 
\end{figure}

For adiabatic fluctuations, the photons are initially perturbed by
$\delta_\gamma(0) = - 2 \psi(0) = 4\mathcal{R}(0)/3$,
i.e.\ they are over-dense in potential
wells, and their velocity vanishes $v_\gamma(0)=0$. If we consider super-Hubble
scales at last scattering, there has been insufficient time for
$v_\gamma$ to grow by gravitational
infall and the action of pressure gradients and it remains small. The
photon continuity equation (\ref{adc:eq22}) then implies that
$\delta_\gamma - 4 \phi$ remains constant, and the decay of $\phi$ through
the matter--radiation transition leaves $(\delta_\gamma/4 + \psi)(\eta_*)
\approx \phi(\eta_\ast)/3 = - 3\mathcal{R}(0)/5$ on large scales ($k <
3\times 10^{-3}\,\mathrm{Mpc}^{-1}$) at last scattering. The combination
$\delta_\gamma/4 +\psi = \Theta_0 + \psi$ is the dominant contribution
to the large-scale temperature anisotropies produces at last scattering;
see (\ref{adc:eq14}).
The evolution of the photon density and velocity
perturbations for adiabatic initial conditions are show in Fig.~\ref{adc:fig3},
along with the scale dependence of the fluctuations at last scattering.
The plateau in $(\delta_\gamma/4 + \psi)(\eta_*)$ on large scales
ensures that a scale-invariant spectrum of curvature perturbations
translates into a scale-invariant spectrum of temperature anisotropies,
$l(l+1)C_l = \mathrm{constant}$, for small $l$.

\begin{figure}[t!]
\begin{center}
\includegraphics[height=8cm,angle=-90]{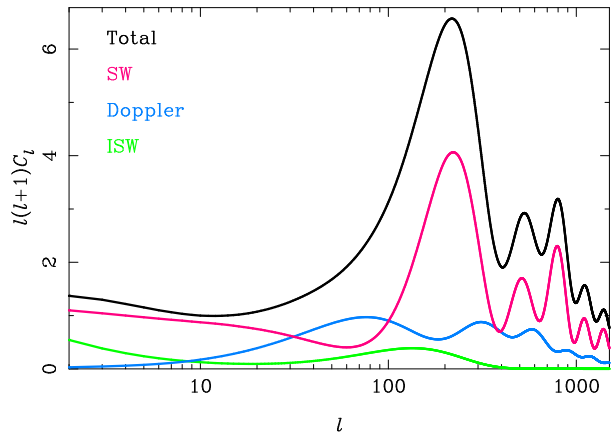}
\end{center}
\caption{Contribution of the various terms in (\ref{adc:eq14}) to the
temperature-anisotropy power spectrum from adiabatic initial conditions.
At high $l$, the contributions
are (from top to bottom): total power; $\delta_\gamma/4 + \psi$ (denoted
SW for Sachs--Wolfe~\cite{adc:sw67}); Doppler effect from $v_\mathrm{b}$;
and the integrated Sachs--Wolfe effect (ISW)
coming from evolution of the potential along the line of sight.
}
\label{adc:fig4} 
\end{figure}

On scales below the sound horizon at last scattering, the photon--baryon
fluid has had time to undergo acoustic oscillation. The form of the photon
initial condition, and the observation that the driving term in
(\ref{adc:eq27}) mimics the cosine WKB solution of the homogeneous equation
(see Fig.~\ref{adc:fig2}), set the oscillation mostly in the
$\cos kr_\mathrm{s}$ mode. The midpoint of the oscillation is roughly at
$\delta_\gamma / 4 = - (1+R)\psi$. This behaviour is illustrated in
Fig.~\ref{adc:fig3}. Modes with $kr_\mathrm{s}(\eta_*)=\pi$ have undergone
half an oscillation at last scattering, and are maximally
compressed. The large value of $\Theta_0 + \psi$ at this particular scale
gives rise to the first acoustic peak in Fig.~\ref{adc:fig1}, now measured
to be at $l=220.1\pm 0.8$~\cite{adc:page03}.
The subsequent extrema of the acoustic oscillation
at $kr_\mathrm{s}(\eta_*)=n\pi$ give rise to the further acoustic peaks. The
angular spacing of the peaks is almost constant and is set by the sound
horizon at last scattering and the angular diameter distance to last
scattering. The acoustic part of the anisotropy spectrum thus encodes a
wealth of information on the cosmological parameters; see
Sect.~\ref{adc:sec:params}. The photon velocity $v_\gamma$ oscillates as
$\sin kr_\mathrm{s}$, so the Doppler term in (\ref{adc:eq14}) tends to fill
in power between the acoustic peaks. The relative phase of the oscillation
of the photon velocity has important implications for the polarization
properties of the CMB as discussed in Sect.~\ref{adc:sec:pol}. The
contributions of the various terms in (\ref{adc:eq14}) to the
temperature-anisotropy power spectrum are shown in
Fig.~\ref{adc:fig4} for adiabatic perturbations.

\subsubsection{Isocurvature Fluctuations}

\begin{figure}[t!]
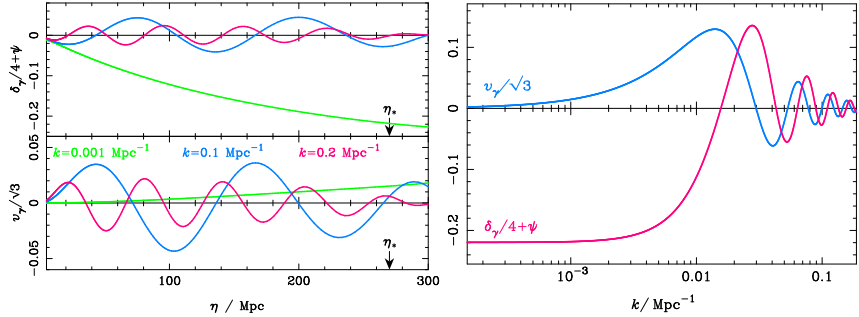

\begin{center}
\includegraphics[width=0.35\textwidth,angle=-90]{adc_fig5a.ps}
\includegraphics[width=0.35\textwidth,angle=-90]{adc_fig5b.ps}
\end{center}
\caption{As Fig.~\ref{adc:fig3} but for CDM-isocurvature initial conditions.
}
\label{adc:fig5} 
\end{figure}

For the CDM-isocurvature mode\footnote{It is also possible to have the
dominant fractional fluctuation in the baryon density rather than the cold
dark matter. However, this mode is nearly indistinguishable from the
CDM mode since, in the absence of baryon pressure, they differ only by a
constant mode in which the radiation and the geometry remain unperturbed, but
the CDM and baryon densities have compensating density
fluctuations~\cite{adc:gordon03}.} the photons are initially unperturbed, as
is the geometry: $\delta_\gamma(0)=0=\phi(0)$ and $v_\gamma = 0$.
On large scales
$\delta_\gamma/4 = \phi$ is preserved, so the growth in $\phi$ during
radiation domination is matched by a growth in $\delta_\gamma$ and the
photons are under-dense in potential wells. It follows that at last scattering
$(\delta_\gamma/4 + \psi)(\eta_*) \approx 2 \phi(\eta_*)$ for $k < 3 \times
10^{-3}\, \mathrm{Mpc}^{-1}$. Note that the redshift climbing out of a
potential well \emph{enhances} the intrinsic temperature fluctuation due to
the photon under-density there. The evolution of the photon fluctuations
for isocurvature initial conditions are shown in Fig.~\ref{adc:fig5}.

\begin{figure}[t!]
\begin{center}
\includegraphics[height=8cm,angle=-90]{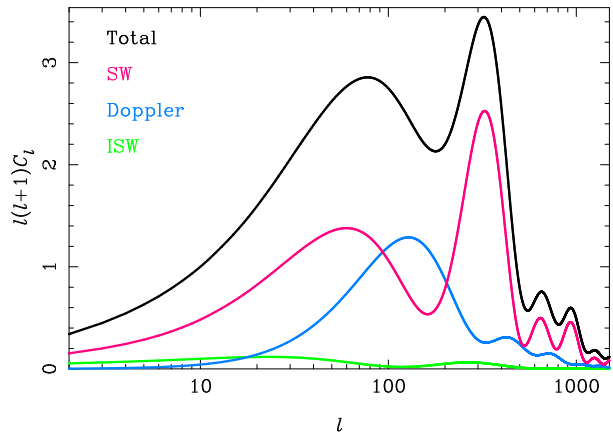}
\end{center}
\caption{As Fig.~\ref{adc:fig4} but for CDM-isocurvature initial conditions.
The initial spectrum of entropy perturbations is scale-invariant.
}
\label{adc:fig6} 
\end{figure}

The evolution of the potential for isocurvature modes makes the driving term
in (\ref{adc:eq27}) mimic the sine solution of the homogeneous equation, and
so $\delta_\gamma$ follows suit oscillating as $\sin \sim kr_\mathrm{s}$ about
the equilibrium point $-4(1+R)\psi$. The
acoustic peaks are at $k r_\mathrm{s}(\eta_*) \sim n\pi/2$, and the
photons are under-dense in the potential wells for the odd-$n$ peaks, while
over-dense in the even $n$. The various contributions to the
temperature-anisotropy power spectrum for isocurvature initial conditions
are shown in Fig.~\ref{adc:fig6}. The different peak positions for isocurvature
initial conditions allow the CMB to constrain their relative contribution
to the total fluctuations. Current constraints are rather dependent on whether
one allows for correlations between the adiabatic and isocurvature modes
(as are generic in the multi-field inflation models that might have generated
the initial conditions), and the extent to which additional cosmological
constraints are employed; see~\cite{adc:bucher04} for a recent analysis
allowing for the most general correlations but a single power-law spectrum.

\subsubsection{Beyond Tight-Coupling}

On small scales it is necessary to go beyond tight-coupling of the
photon--baryon system since the photon diffusion length can become comparable
to the wavelength of the fluctuations. Photons that have had sufficient time
to diffuse of the order of a wavelength can leak out of over-densities, thus
damping the acoustic oscillations and generating anisotropy~\cite{adc:silk68}.
A rough estimate of the comoving scale below which diffusion is important is
the square root of the geometric mean of the particle horizon (or conformal
age) and the mean-free path of the photons, i.e.\ $\sqrt{\eta / |\dot{\tau}|}$.
Converting this to a comoving wavenumber defines the damping scale
\begin{equation}
k_\mathrm{D}^{-2} \sim 0.3 (\Omega_\mathrm{m}h^2)^{-1/2}
(\Omega_\mathrm{b}h^2)^{-1} (a/a_*)^{5/2}\, \mathrm{Mpc}^2
\label{adc:eq31}
\end{equation}
when the scale factor is $a$. Here, $a_*$ is the scale factor at last
scattering, and the expression is valid well after matter--radiation equality
but well before recombination. The effect of diffusion is to damp the
photon (and baryon) oscillations exponentially by the time of last scattering
on comoving scales smaller than $\sim 3\,\mathrm{Mpc}$. The resulting damping
effect on the temperature power spectrum has now been measured by several
experiments~\cite{adc:dickinson04,adc:pearson03,adc:kuo04}.

To describe diffusion damping more quantitatively, we consider 
scales that were already sub-Hubble during radiation domination. The
gravitational potentials will then have been suppressed during their
oscillatory phase when the photons (which are undergoing acoustic oscillations
themselves) dominated the energy density, and so we can ignore gravitational
effects.
Furthermore, the dynamical timescale of the acoustic oscillations is then
short compared to the expansion time and we can ignore the effects of
expansion. In this limit, the Euler equations for the photons and the baryons
can be iterated to give the relative velocity between the photons and
baryons to first order in $k/|\dot{\tau}|$:
\begin{equation}
(1+ R^{-1}) (v_\gamma - v_\mathrm{b}) = \frac{1}{4}
k \dot{\tau}^{-1} \delta_\gamma\; .
\label{adc:eq32}
\end{equation}
Using momentum conservation for the total photon--baryon system gives
\begin{equation}
\dot{v}_\gamma + R \dot{v}_\mathrm{b} + \frac{1}{4}
k \delta_\gamma - \frac{2}{5} k\Theta_2 = 0\; ,
\label{adc:eq33}
\end{equation}
which can be combined with the derivative of (\ref{adc:eq32}) to give a new
Euler equation for the photons correct to first order in tight coupling:
\begin{equation}
(1+R)\dot{v}_\gamma \approx - \frac{1}{4} k\delta_\gamma + \frac{R^2}{4(1+R)}
k \dot{\tau}^{-1} \dot{\delta}_\gamma + \frac{16}{45} k^2 \dot{\tau}^{-1}
v_\gamma\; .
\label{adc:eq34}
\end{equation}
Here, we have used $\Theta_2 \approx 8 k \dot{\tau}^{-1} v_\gamma /9$
which includes the correction due to polarization.
In the limit of perfect coupling, (\ref{adc:eq34}) reduces to
(\ref{adc:eq26}) on small scales. The continuity equation for the photons,
$\dot{\delta}_\gamma = 4 k v_\gamma /3$ ($+ 4\dot{\phi}$), shows that
the last two terms on the right of (\ref{adc:eq34}) are drag terms, and on
differentiating gives
\begin{equation}
\ddot{\delta}_\gamma - \frac{k^2 \dot{\tau}^{-1}}{3(1+R)}\left(
\frac{16}{15} + \frac{R^2}{1+R}\right)\dot{\delta}_\gamma + \frac{k^2}{3(1+R)}
\delta_\gamma = 0\; .
\label{adc:eq35}
\end{equation}
The WKB solution is
\begin{equation}
\delta_\gamma \propto \E^{\pm \I k r_\mathrm{s}}
\E^{-k^2/k_\mathrm{D}^2}\; ,\quad \mathrm{where} \quad \frac{1}{k_\mathrm{D}^2}
\equiv \frac{1}{6} \int_0^\eta
\frac{|\dot{\tau}^{-1}|}{1+R}\left(\frac{16}{15}+
\frac{R^2}{1+R}\right) \, \D\eta'
\label{adc:eq36}
\end{equation}
is the damping scale.

The finite mean-free path of CMB photons around last scattering has an
additional effect on the temperature anisotropies. The visibility function
$-\dot{\tau}\E^{\tau}$ has a finite width $\sim 80\, \mathrm{Mpc}$ and so along
a given line of sight photons will be last scattered over this interval.
Averaging over scattering events will tend to wash out the anisotropy
from wavelengths short compared to the width of the visibility function.
This effect is described mathematically by integrating the oscillations in the
spherical Bessel functions in (\ref{adc:eq17}) against the product of the
visibility function and the (damped) perturbations.

Boltzmann codes such
as CMBFAST~\cite{adc:seljak96} and CAMB~\cite{adc:lewis00}
use the tight-coupling
approximation at early times to avoid the numerical problems associated with
integrating
the stiff Euler equations in their original forms (\ref{adc:eq23}) and
(\ref{adc:eq24}).

\subsection{Other Features of the Temperature-Anisotropy Power Spectrum}
\label{adc:subsec:other}

We end this section on the fundamentals of the physics of CMB temperature
anisotropies by reviewing three additional effects that contribute to the
linear anisotropies.

\subsubsection{Integrated Sachs--Wolfe Effect}
\label{adc:subsec:ISW}

The integrated Sachs--Wolfe (ISW) effect is described by the last term on
the right of (\ref{adc:eq14}). It is an additional source of anisotropy due
to the temporal variation of the gravitational potentials along the line of
sight: if a potential well deepens as a CMB photon crosses it then the
blueshift due to infall will be smaller than redshift from climbing out of the
(now deeper) well.
(The combination $\phi + \psi$ has a direct geometric interpretation
as the potential for the electric part of the Weyl
tensor~\cite{adc:stewart90}.)
The ISW receives contributions from late times as the potentials decay
during dark-energy domination, and at early times around last scattering
due to the finite time since matter--radiation equality.

The late-time effect
contributes mainly on large angular scales since there is little power
in the potentials at late times on scales that entered the Hubble radius
during radiation domination. The late ISW effect is the only way to probe
late-time structure growth (and hence e.g.\ distinguish between different
dark-energy models) with linear CMB anisotropies, but this is hampered by
cosmic variance on large angular scales. The late ISW effect produces
correlations between the large-scale temperature fluctuations and
other tracers of the potential in the local universe, and with the advent of
the WMAP data these have now been
tentatively detected~\cite{adc:boughn04,adc:nolta03,adc:fosalba03};
see also Sect.~\ref{adc:sec:highlights}.

In adiabatic models the early-time ISW effect adds coherently with the
contribution $\delta_\gamma/4 + \psi$ to the anisotropies near the first peak,
boosting this peak significantly~\cite{adc:hu95a}; see Fig.~\ref{adc:fig4}.
The reason is that the linear scales that
contribute here are maximally compressed with $\delta_\gamma/4 + \psi \sim
- \psi / 2$ which has the same sign as $\dot{\phi}$ for decaying $\phi$.

\subsubsection{Reionization}
\label{adc:subsec:reionization}

\begin{figure}[t!]
\begin{center}
\includegraphics[height=8cm,angle=-90]{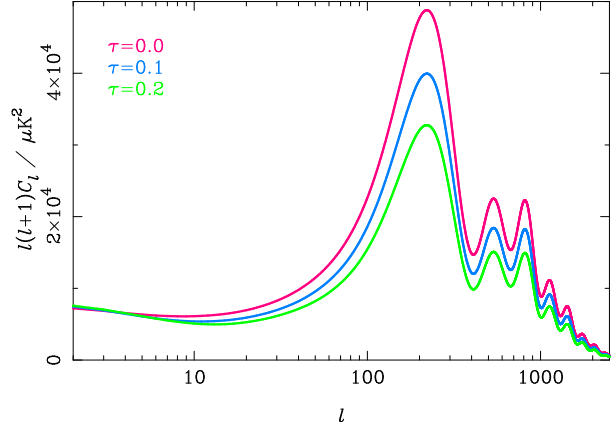}
\end{center}
\caption{Effect of reionization on the temperature-anisotropy power spectrum.
The spectra are (from top to bottom) for no reionization,
$\tau_\mathrm{re}=0.1$ and $0.2$.
}
\label{adc:fig7} 
\end{figure}

Once structure formation had proceeded to produce the first sources of
ultra-violet photons, the universe began to reionize. The resulting free
electron density could then re-scatter CMB photons, and this tended to
isotropise the CMB by averaging the anisotropies from many lines of sight at
the scattering event. Approximating the bi-modal visibility function
as two delta functions, one at last scattering\footnote{We continue to
refer to the last scattering event around recombination as last scattering,
even in the presence of re-scattering at reionization.} $\eta_*$ and one at
reionization $\eta_\mathrm{re}$, if the optical depth through reionization
is $\tau_\mathrm{re}$, the temperature fluctuation at $\vx=0$ at $\eta_0$ is
\begin{eqnarray}
[\Theta(\vnhat) + \psi]_{\eta_0} &\approx& (1-\E^{-\tau_\mathrm{re}})
(\Theta_0 + \psi - \vnhat \cdot \vvb)[-\vnhat(\eta_0-\eta_\mathrm{re}),
\eta_\mathrm{re}] \nonumber \\
&&\mbox{}+ \E^{-\tau_\mathrm{re}}(\Theta_0 +\psi - \vnhat\cdot \vvb)
[-\vnhat(\eta_0-\eta_*),\eta_*]\; .
\label{adc:eq37}
\end{eqnarray}
Here, we have used (\ref{adc:eq13}), neglected the ISW effect, and approximated
the scattering as isotropic. The first term on the right describes the
effect of blending the anisotropies from different lines of sight (to give
$\Theta_0$) and the generation of new anisotropies by re-scattering off moving
electrons at reionization; the second term is simply the temperature anisotropy
that would be observed with no reionization, weighted by the fraction of
photons that do not re-scatter.
Since $\Theta_0 + \psi$ at the re-scattering event
is the average of $\Theta_0 + \psi - \vnhat'\cdot \vvb$ on the electron's
last scattering surface, on large scales $k (\eta_\mathrm{re}-\eta_*) \ll 1$
it reduces to $\Theta_0 + \psi$ at $[-\vnhat(\eta_0-\eta_*),\eta_*]$, while
on small scales it vanishes. It follows that for scales that are super-horizon
at reionization, the observed temperature anisotropy becomes
\begin{equation}
\Theta(\vnhat) \rightarrow \Theta(\vnhat) - (1-\E^{-\tau_\mathrm{re}})
\vnhat \cdot \Delta \vvb \;,
\label{adc:eq38}
\end{equation}
where $\Delta \vvb$ is the difference between the electron velocity at
the reionization event and the preceding last scattering event on
the line of sight. On such scales the Doppler terms do not contribute
significantly and the temperature anisotropy is unchanged. For scales that
are sub-horizon at reionization,
\begin{equation}
\Theta(\vnhat) \rightarrow \E^{-\tau_\mathrm{re}}\Theta(\vnhat) -
(1-\E^{-\tau_\mathrm{re}}) \vnhat \cdot \vvb \;,
\label{adc:eq39}
\end{equation}
where the Doppler term is evaluated at reionization. In practice, the
visibility function is not perfectly sharp at reionization and the integral
through the finite re-scattering distance tends to wash out the Doppler term
since only plane waves with their wavevectors near the line of sight
contribute significantly to $\vnhat \cdot \vvb$.
Figure~\ref{adc:fig7} shows the resulting effect
$C_l \rightarrow \E^{-2\tau_\mathrm{re}} C_l$ on the anisotropy power
spectrum on small scales. Recent results from WMAP~\cite{adc:kogut03} suggest
an optical depth through reionization $\tau_\mathrm{re}\sim 0.17$.
Such early reionization cannot have
been an abrupt process since the implied redshift $z_\mathrm{re} \sim 15$
is at odds with the detection of traces of smoothly-distributed
neutral hydrogen at $z\sim 6$ via Gunn-Peterson troughs in the spectra of
high-redshift quasars~\cite{adc:becker01,adc:djorgovski01}.

\subsubsection{Tensor Modes}
\label{adc:subsec:tensors}

Tensor modes, describing gravitational waves, represent the
transverse trace-free perturbations to the spatial metric:
\begin{equation}
\D s^2 = a^2(\eta) [\D \eta^2 - (\delta_{ij} + h_{ij})\D x^i \D x^j]\; , 
\label{adc:eq40}
\end{equation}
with $h_i^i=0$ and $\partial_i h^i_j = 0$. A convenient parameterisation of
the photon four-momentum in this case is
\begin{equation}
p^\mu = \frac{\epsilon}{a^2}\left[1, e^i - \frac{1}{2} h^i_j e^j\right]\; ,
\label{adc:eq41}
\end{equation}
where $\ve^2=1$ and $\epsilon$ is $a$ times the energy of the photon as seen
by an observer at constant $\vx$. The components of $\ve$ are the projections
of the photon direction for this observer on an orthonormal
spatial triad of vectors $a^{-1}(\partial_i - h_i^j \partial_j / 2)$.
In the background $\ve = \D \vx / \D \eta$ and is constant. The evolution of
the comoving energy $\epsilon$ in the perturbed universe is
\begin{equation}
\frac{1}{\epsilon} \frac{\D \epsilon}{\D \eta} + \frac{1}{2} \dot{h}_{ij}
e^i e^j = 0\; ,
\label{adc:eq42}
\end{equation}
and so the Boltzmann equation for $\Theta(\eta,\vx,\ve)$ is
\begin{eqnarray}
\frac{\partial \Theta}{\partial \eta} + \ve \cdot \vec{\nabla}
\Theta  &=& - a \nel \sigT \Theta + \frac{3}{16\pi}
a \nel\sigT \int \Theta(\ve')[(1+(\ve\cdot\ve')^2]\, \D\ve' \nonumber \\
&&\mbox{} - \frac{1}{2} \dot{h}_{ij} e^i e^j\; .
\label{adc:eq43}
\end{eqnarray}
Neglecting the anisotropic nature of Thomson scattering, the solution of this
equation is an integral along the unperturbed line of sight:
\begin{equation}
\Theta(\vnhat) = - \frac{1}{2} \int_0^{\eta_0}\E^{-\tau} \dot{h}_{ij} \hat{n}^i
\hat{n}^j \, \D\eta\; .
\label{adc:eq44}
\end{equation}
The time derivative $\dot{h}_{ij}$ is the shear induced by the gravitational
waves. This quadrupole perturbation to the expansion produces an anisotropic
redshifting of the CMB photons and an associated temperature anisotropy.

\begin{figure}[t!]
\begin{center}
\includegraphics[height=8cm,angle=-90]{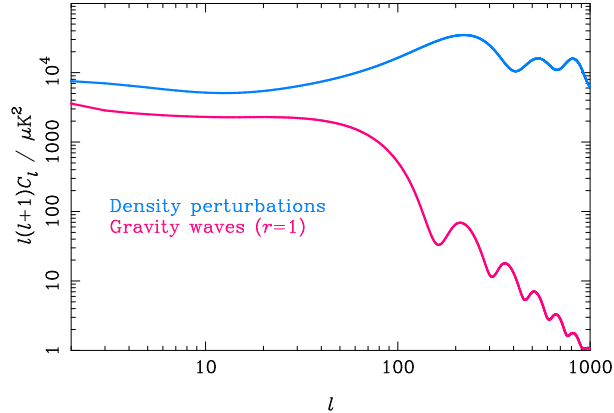}
\end{center}
\caption{The temperature-anisotropy power spectrum from scalar
perturbations (density perturbations; top) and tensor perturbations (gravity
waves; bottom) for a tensor-to-scalar ratio $r=1$.
}
\label{adc:fig8} 
\end{figure}

Figure~\ref{adc:fig8} compares the power spectrum due to gravitational
waves with that from scalar perturbations for a tensor-to-scalar ratio
$r=1$ corresponding to an energy scale of inflation
$3.3\times 10^{16}\,\mathrm{GeV}$.
The constraints on gravitational waves from temperature anisotropies are
not very constraining since their effect is limited to large angular scales
where cosmic variance from the dominant scalar perturbations is large.
Gravitational waves damp as they oscillate
inside the horizon, so the only significant anisotropies are from wavelengths
that are super-horizon at last scattering, corresponding to $l \sim 60$.
The current 95-per cent upper limit on the tensor-to-scalar ratio is
0.68~\cite{adc:dickinson04}.
Fortunately, CMB polarization provides an alternative route to
detecting the effect of gravitational waves on the CMB which is not
limited by cosmic variance~\cite{adc:seljak97,adc:kamionkowski97};
see also Sect.~\ref{adc:sec:pol}.

\section{Cosmological Parameters and the CMB}
\label{adc:sec:params}

The simple, linear physics of CMB temperature anisotropies, reviewed in the
previous section, means that the CMB depends sensitively on many of the
key cosmological parameters. For this reason, CMB observations over the past
decade have been a significant driving force in the quest for precision
determinations of the cosmological parameters. It is not our intention here
to give a detailed description of the constraints that have emerged from
such analyses, e.g.~\cite{adc:bond03},
but rather to provide a brief description
of how the key parameters affect the temperature-anisotropy power spectrum.
More details can be found in the seminal papers on this subject,
e.g.~\cite{adc:hu95a,adc:hu95b,adc:bond94}
and references therein.

\subsection{Matter and Baryons}

\begin{figure}[t!]
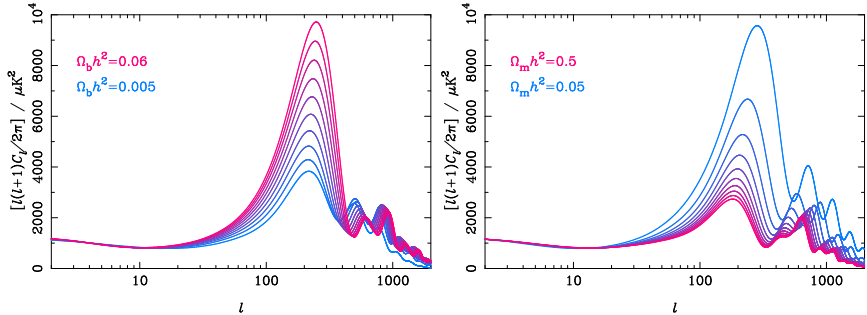

\begin{center}
\includegraphics[width=0.35\textwidth,angle=-90]{adc_fig9a.ps}
\includegraphics[width=0.35\textwidth,angle=-90]{adc_fig9b.ps}
\end{center}
\caption{Dependence of the temperature-anisotropy power spectrum on the
physical density in baryons (left) and all non-relativistic matter (right).
From top to bottom at the first peak, the baryon densities vary linearly
in the range $\Omega_\mathrm{b}h^2 = 0.06$--0.005 (left) and the
matter densities in $\Omega_\mathrm{m}h^2=0.05$--$0.5$ (right). The
initial conditions are adiabatic.
}
\label{adc:fig9} 
\end{figure}

The curvature of the universe and the properties of the dark energy are
largely irrelevant for the pre-recombination physics of the acoustic
oscillations. Their main contribution is felt geometrically through
the angular diameter distance to last scattering, $D_\mathrm{A}$, which
controls the projection of linear scales there to angular scales on the sky
today. In contrast, those parameters that determine the energy content
of the universe before recombination, such as the
physical densities in (non-relativistic) matter $\Omega_\mathrm{m}h^2$,
and radiation $\Omega_\mathrm{r}h^2$ (determined by the CMB temperature and
the physics of neutrinos), play an important role in acoustic physics by
determining the expansion rate and hence the behaviour of the perturbations.
In addition, the physical density in baryons, $\Omega_\mathrm{b}h^2$, affects
the acoustic oscillations through baryon inertia and the dependence of the
photon mean-free path on the electron density. The effect of
variations in the physical densities of the matter and baryon densities
on the anisotropy power spectrum is illustrated in Fig.~\ref{adc:fig9}
for adiabatic initial conditions.

The linear scales at last scattering that have reached
extrema of their oscillation are determined by the initial conditions (i.e.\
adiabatic or isocurvature) and the sound horizon $r_\mathrm{s}(\eta_*)$.
Increasing the baryon density holding the total matter density fixed reduces
the sound speed while preserving the expansion rate (and moves
last scattering to slightly earlier times). The effect is to reduce
the sound horizon at last scattering and so the wavelength of those modes
that are at extrema of their oscillation, and hence push the acoustic
peaks to smaller scales. This effect could be confused with a change
in the angular diameter distance $D_\mathrm{A}$, but fortunately baryons
have another distinguishing effect. Their inertia shifts the zero point of
the acoustic oscillations to $\sim -(1+R)\psi$, and enhances the amplitude
of the oscillations. In adiabatic models for modes that enter the sound
horizon in matter domination, $\delta_\gamma/4$ starts out at $-2\psi/3$,
and so the amplitude of the oscillation is $-\psi(1+3R)/3$. The combination
of these two effects is to enhance the amplitude of $\Theta_0 +\psi$ at
maximal compression by a factor of $1+6R$ over that at minimal compression.
The effect on the power spectrum is to enhance the amplitude of the 1st, 3rd
etc.\ peaks for adiabatic initial conditions, and the 2nd, 4th etc.\ for
isocurvature. Current CMB data gives $\Omega_\mathrm{b}h^2 = 0.023 \pm 0.001$
for power-law $\Lambda$CDM models~\cite{adc:spergel03},
beautifully consistent with
determinations from big bang nucleosynthesis. Other effects of baryons
are felt in the damping tail of the power spectrum since increasing
the baryon density tends to inhibit diffusion giving less damping at a given
scale.

The effect of increasing the physical matter density $\Omega_\mathrm{m}h^2$
at fixed $\Omega_\mathrm{b} h^2$ is also two-fold (see Fig.~\ref{adc:fig9}):
(i) a shift of the peak
positions to larger scales due to the increase in $D_\mathrm{A}$; and (ii)
a scale-dependent reduction in peak height in adiabatic models. Adiabatic
modes that enter the sound horizon during radiation domination see the
potentials decay as the photon density rises to reach maximal compression.
This decay tends to drive the oscillation, increasing the oscillation
amplitude. Raising $\Omega_\mathrm{m}h^2$ brings matter--radiation
equality to earlier times, and reduces the efficiency of the gravitational
driving effect for the low-order peaks. Current CMB data gives
$\Omega_\mathrm{m}h^2 = 0.13 \pm 0.01$ for adiabatic, power-law $\Lambda$CDM
models~\cite{adc:spergel03}.

\subsection{Curvature, Dark Energy and Degeneracies}
\label{adc:subsec:curvature}

\begin{figure}[t!]
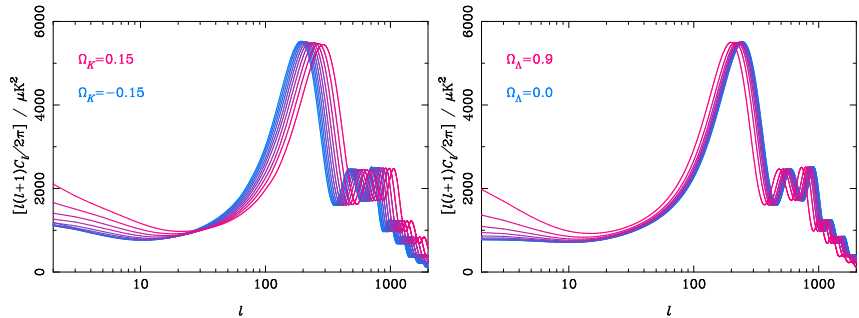

\begin{center}
\includegraphics[width=0.35\textwidth,angle=-90]{adc_fig10a.ps}
\includegraphics[width=0.35\textwidth,angle=-90]{adc_fig10b.ps}
\end{center}
\caption{Dependence of the temperature-anisotropy power spectrum on
the curvature $\Omega_K$ (left) and cosmological constant $\Omega_\Lambda$
(right) in adiabatic models. In both cases, the physical densities
in baryons and matter were held constant, thus preserving the conditions on the
last scattering surface. The curvature varies (left to right) in the
range -0.15--0.15 and the cosmological constant in the range 0.9--0.0.
}
\label{adc:fig10} 
\end{figure}

The main effect of curvature and dark energy on the linear CMB anisotropies
is through the angular diameter distance and the late-time integrated
Sachs--Wolfe effect; see Fig.~\ref{adc:fig10} for the case of adiabatic
fluctuations in cosmological-constant models. The ISW contribution is
limited to large scales where cosmic variance severely limits the precision
of power spectrum estimates. There is an additional small effect due to
quantisation of the allowed spatial modes in closed
models (e.g.~\cite{adc:abbott86}), but
this is also confined to large scales
(i.e.\ near the angular projection of the curvature
scale). Most of the information that the CMB encodes on curvature and
dark energy is thus locked in the angular diameter distance to last
scattering, $D_\mathrm{A}$.

\begin{figure}[t!]
\begin{center}
\includegraphics[height=8cm,angle=-90]{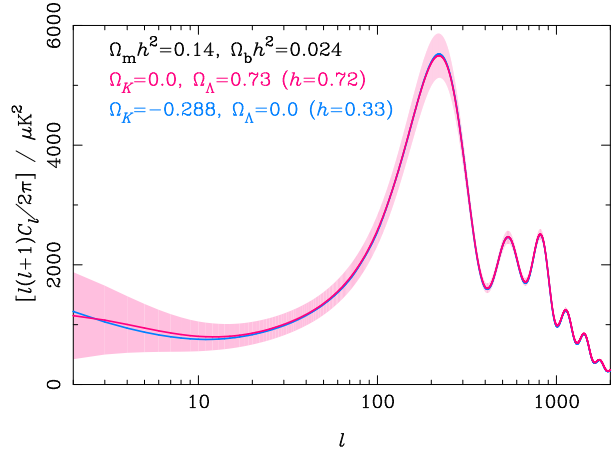}
\end{center}
\caption{The geometric degeneracy. A scale-invariant adiabatic
$\Lambda$CDM model with
$\Omega_\mathrm{b}h^2 = 0.024$, $\Omega_\mathrm{m}h^2 = 0.14$ and
$\Omega_\Lambda = 0.73$ and $\Omega_K = 0$ (close to the WMAP best-fit
values~\cite{adc:spergel03})
produces an almost identical spectrum to a closed model
$\Omega_K=-0.288$ with vanishing cosmological constant. However, the
Hubble constants are very different -- $h=0.72$ in the flat model and
$0.33$ in the closed model -- and so the latter is easily ruled out by
external constraints. The shaded region shows the $1\sigma$ cosmic variance
errors $\Delta C_l/C_l = \sqrt{2/(2l+1)}$ on the power spectrum.
}
\label{adc:fig11} 
\end{figure}

With the physical densities
$\Omega_\mathrm{b}h^2$ and $\Omega_\mathrm{m}h^2$ fixed by the acoustic part
of the anisotropy spectrum, $D_\mathrm{A}$ can be considered a function of
$\Omega_K$ and the history of the energy density of the dark energy (often
modelled through its current density and a constant equation of state). In
cosmological
constant models $D_A$ is particularly sensitive to the curvature: the
95-per cent interval from WMAP alone (with the weak prior $H_0 > 50\,
\mathrm{km}\,\mathrm{s}^{-1}\,\mathrm{Mpc}^{-1}$) is
$-0.08 < \Omega_K < 0.02$, so the universe is
close to being spatially flat. The fact that the impact
of curvature and the properties of the dark energy on the CMB is mainly
through a single number $D_\mathrm{A}$ leads to a geometrical degeneracy
in parameter estimation~\cite{adc:efstathiou99},
as illustrated in Fig.~\ref{adc:fig11}.
Fortunately, this is easily broken by including other, complementary
cosmological datasets. The constraint on curvature from WMAP improves
considerably when supernovae
measurements~\cite{adc:reiss98,adc:perlmutter99},
or the measurement of $H_0$ from the Hubble Space Telescope Key
Project~\cite{adc:freedman01} are included.
Other examples of near-perfect degeneracies for
the temperature anisotropies include the addition of gravity waves and
a reduction in the amplitude of the initial fluctuations mimicing the
effect of reionization. This degeneracy is broken very effectively by
the polarization of the CMB.

\section{CMB Polarization}
\label{adc:sec:pol}

The growth in the mean-free path of the CMB photons during recombination
allowed anisotropies to start to develop. Subsequent scattering of the
radiation generated (partial) linear polarization from the quadrupole
anisotropy. This linear polarization signal is expected to have an
r.m.s.\ $\sim 5\, \mu\mathrm{K}$, and, for scalar perturbations, to peak
around multipoles $l \sim 1000$ corresponding to the angle subtended by the
mean-free path around last scattering. The detection of CMB polarization
was first announced in 2002 by the Degree Angular Scale Interferometer (DASI)
team~\cite{adc:kovac02}; WMAP has also detected the polarization
indirectly through
its correlation with the temperature anisotropies~\cite{adc:kogut03}. A direct
measurement of the polarization power from two-years of WMAP data is expected
shortly. Polarization is only generated by scattering, and so is a sensitive
probe of conditions at recombination. In addition, large-angle polarization
was generated by subsequent re-scattering as the universe reionized, providing
a unique probe of the ionization history at high redshift.

\subsection{Polarization Observables}
\label{adc:subsec:pol_obs}

Polarization is conveniently described in terms of Stokes parameters
$I$, $Q$, $U$ and $V$, where $I$ is the total intensity discussed at length
in the previous section. The parameter $V$ describes circular polarization
and is expected to be zero for the CMB since it is not generated by Thomson
scattering. The remaining parameters $Q$ and $U$ describe linear polarization.
They are the components of the trace-free, (zero-lag) correlation tensor of the
electric field in the radiation, so that for a quasi-monochromatic
plane wave propagating along the $z$ direction
\begin{equation}
\left( \begin{array}{cc}
	\langle E_x^2 - E_y^2 \rangle  & 2\langle E_x E_y \rangle \\
	2\langle E_x E_y \rangle	& - \langle E_x^2 - E_y^2 \rangle
	\end{array}
\right) = \frac{1}{2} \left(
\begin{array}{cc}
Q & U \\
U & -Q
\end{array}
\right)\; ,	
\label{adc:eq45}
\end{equation}
where the angle brackets represent an average on timescales
long compared to the period of the wave. For diffuse radiation we
define the polarization brightness tensor $\mathcal{P}_{ab}(\vnhat)$ to have
components given by (\ref{adc:eq45}) for plane waves within a bundle around
the line of sight $\vnhat$ and around the specified frequency. The polarization
tensor is transverse to the line of sight, and, since it
inherits its frequency dependence from the
the quadrupole of the total intensity, has a spectrum given by
the derivative of the Planck function (see equation~\ref{adc:eq11}).  

\begin{figure}[t!]
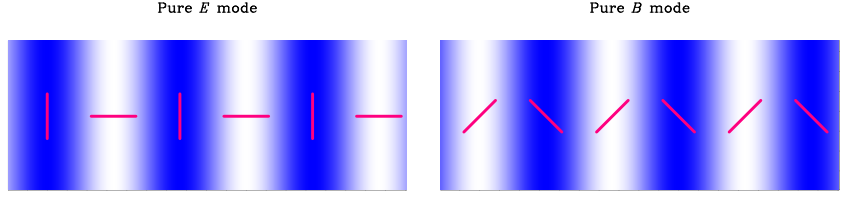

\begin{center}
\includegraphics[angle=-90,width=0.45\textwidth]{adc_fig12a.ps}
\quad
\includegraphics[angle=-90,width=0.45\textwidth]{adc_fig12b.ps}
\end{center}
\caption{Polarization patterns for a pure-electric mode (left) and
pure-magnetic mode (right) on a small patch of the sky for potentials that
are locally Fourier modes. The shading denotes the amplitude of the potential.
For the electric pattern the polarization is aligned with or perpendicular to
the Fourier wavevector depending on the sign of the potential; for the
magnetic pattern the polarization is at 45 degrees. 
}
\label{adc:fig12} 
\end{figure}

The polarization tensor can be decomposed uniquely on the sphere into
an electric (or gradient) part and a magnetic (or curl)
part~\cite{adc:seljak97,adc:kamionkowski97}:
\begin{equation}
\mathcal{P}_{ab} = \nabla_{\langle a} \nabla_{b\rangle} P_E -
\epsilon^c{}_{\langle a}\nabla_{b\rangle} \nabla_c P_B \; ,
\label{adc:eq46}
\end{equation}
where angle brackets denote the symmetric, trace-free part, $\nabla_a$
is the covariant derivative on the sphere, and $\epsilon_{ab}$ is the
alternating tensor. The divergence $\nabla^a \mathcal{P}_{ab}$ is a
pure gradient if the magnetic part $P_B = 0$, and a curl if the electric
part $P_E = 0$. The potential $P_E$ is a scalar under parity, but $P_B$
is a pseudo-scalar. For a given potential $P$, the electric and magnetic
patterns it generates (i.e.\ with $P_E = P$ and $P_B = P$ respectively)
are related by locally rotating the polarization directions by 45 degrees.
The polarization orientations on a small patch of the sky for potentials
that are locally Fourier modes are shown in Fig.~\ref{adc:fig12}.
The potentials can be expanded in spherical
harmonics (only the $l\geq 2$ multipoles contribute to $\mathcal{P}_{ab}$)
as
\begin{equation}
P_E(\vnhat) = \sum_{lm} \sqrt{\frac{(l-2)!}{(l+2)!}} E_{lm} Y_{lm}(\vnhat)\; ,
\quad P_B(\vnhat) = \sum_{lm} \sqrt{\frac{(l-2)!}{(l+2)!}} B_{lm}
Y_{lm}(\vnhat)\;.
\end{equation}
(The normalisation is conventional.)
Under parity $E_{lm} \rightarrow (-1)^l E_{lm}$ but $B_{lm} \rightarrow
- (-1)^l B_{lm}$. 
Assuming rotational and parity invariance, $B$ is not
correlated with $E$ or the temperature anisotropies $T$, leaving four
non-vanishing power spectra: $C_l^T$, $C_l^E$, $C_l^B$ and the
cross-correlation $C_l^{TE}$, where e.g.\ $\langle E_{lm} T_{lm}^\ast \rangle
= C_l^{TE}$.

\subsection{Physics of CMB Polarization}
\label{adc:subsec:pol_physics}

For scalar perturbations, the quadrupole of the temperature anisotropies
at leading order in tight coupling is $\Theta_2 \sim
k \dot{\tau}^{-1} v_\gamma$. Scattering of this quadrupole into the direction
$-\vnhat$ generates linear polarization parallel or perpendicular to the
projection of the
wavevector $\vk$ onto the sky, i.e.\ $\mathcal{P}_{ij} \sim \Theta_2
[\hat{k}_{\langle i} \hat{k}_{j\rangle}]^{\mathrm{TT}}$, where TT denotes the
transverse (to $\vnhat$), trace-free part. In a flat universe the polarization
tensor is conserved in the absence of scattering; for non-flat models this
is still true if the components are defined on an appropriately-propagated
basis (e.g.~\cite{adc:challinor00}).
For a single plane wave perturbation, the polarization on the
sky is thus purely electric (see Fig.~\ref{adc:fig12}). For tensor
perturbations, the polarization $\mathcal{P}_{ij} \sim \dot{\tau}^{-1}
[\dot{h}_{ij}]^{\mathrm{TT}}$ since the tightly-coupled quadrupole is
proportional to the shear $\dot{h}_{ij}$. The gravitational wave defines
additional directions on the sky when its shear is projected, and the
polarization pattern is not purely electric. Thus density
perturbations do not produce magnetic polarization in linear perturbation
theory, while gravitational waves produce both electric and
magnetic~\cite{adc:seljak97,adc:kamionkowski97}.

\begin{figure}[t!]
\begin{center}
\includegraphics[height=10cm,angle=-90]{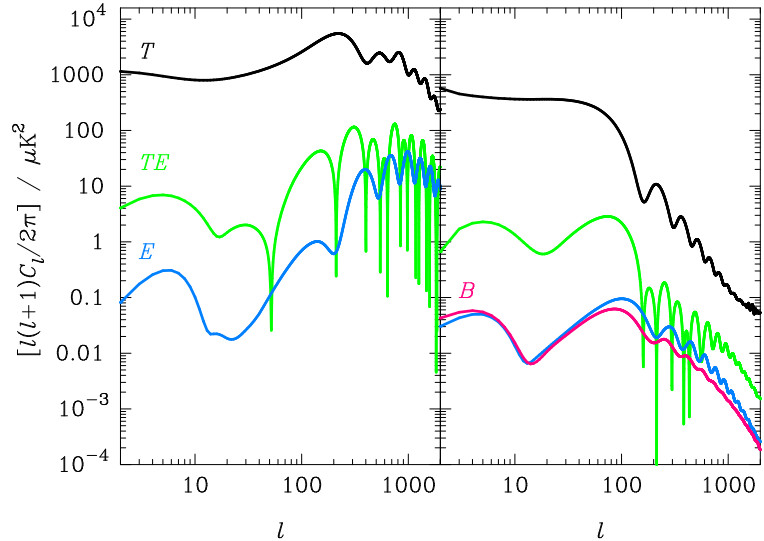}
\end{center}
\caption{Power spectra produced by adiabatic scalar perturbations (left) and
tensor perturbations (right) for a tensor-to-scalar ratio $r=1$. On large
scales the spectra from scalar perturbations are (from top to bottom)
$C_l^T$, $C_l^{TE}$ and $C_l^E$. For tensor perturbations, they are
$C_l^T$, $C_l^{TE}$, $C_l^B$ and $C_l^E$.
}
\label{adc:fig13} 
\end{figure}

The polarization power spectra produced by scalar and tensor perturbations
are compared in Fig.~\ref{adc:fig13}. The scalar $C_l^E$ spectrum peaks around
$l \sim 1000$ since this corresponds to the projection of linear scales at
last scattering for which diffusion generates a radiation quadrupole most
efficiently. The polarization probes the photon bulk velocity at last
scattering, and so $C_l^E$ peaks at the troughs of $C_l^T$, while
$C_l^{TE}$ is zero at the peaks and troughs, and has its extrema
in between. For adiabatic
perturbations, the large-scale cross-correlation changes sign at $l \sim 50$,
and, with the conventions adopted here\footnote{The sign of $E_{lm}$ for
a given polarization field depends on the choice of conventions for
the Stokes parameters and their decomposition into electric and magnetic
multipoles. We follow~\cite{adc:lewis02},
which produces the same sign of $C_l^{TE}$ as~\cite{adc:hu98}, but note that
the Boltzmann codes CMBFAST~\cite{adc:seljak96} and
CAMB~\cite{adc:lewis00}
have the opposite sign.} is positive between $l=50$ and the
first acoustic peak in $C_l^T$. Isocurvature modes produce a negative
correlation from $l=2$ to the first acoustic trough.

Tensor modes produce similar power in electric and magnetic polarization.
As gravitational waves damp inside the horizon, the polarization
peaks just shortward of the horizon size at last scattering $l \sim 100$
despite these large scales being geometrically less efficient at transferring
power to the quadrupole during a mean-free time than smaller scales.

For both scalar and tensor perturbations, the polarization would be small on
large scales were it not for reionization, since a significant quadrupole
is only generated at last scattering when the mean-free path approaches the
wavelength of the fluctuations. However, reionization does produce significant
large-angle polarization~\cite{adc:zaldarriaga97}
(see Fig.~\ref{adc:fig13}). The
temperature quadrupole at last scattering peaks on linear scales with
$k(\eta_{\mathrm{re}}-\eta_*) \sim 2$, which then re-projects onto angular
scales $l \sim 2(\eta_0 - \eta_\mathrm{re})/(\eta_\mathrm{re}-\eta_*)$.
The position of the reionization feature is thus controlled by the
epoch of reionization, and the height by the fraction of photons that
scatter there i.e.\ $\tau_\mathrm{re}$. The measurement of $\tau_\mathrm{re}$
with large-angle polarization allows an accurate determination of the amplitude
of scalar fluctuations from the temperature-anisotropy power spectrum.
In addition, the fine details of the large-angle polarization power
can in principle distinguish different ionization histories with the
same optical depth, although this is hampered by the large cosmic variance
at low $l$~\cite{adc:holder03}.

\section{Highlights of Recent Results}
\label{adc:sec:highlights}

In this section we briefly review some of the highlights from recent
observations of the CMB temperature and polarization anisotropies.
Analysis of the former have entered a new phase with the release of the
first year data from the WMAP satellite~\cite{adc:bennett03};
a further three years
worth of data are expected from this mission. Detections of CMB polarization
are still in their infancy, but here too we can expect significant progress
from a number of experiments in the short term.

\subsection{Detection of CMB Polarization}
\label{adc:subsec:polhighlights}

\begin{figure}[t!]
\begin{center}
\includegraphics[height=8cm,angle=-90]{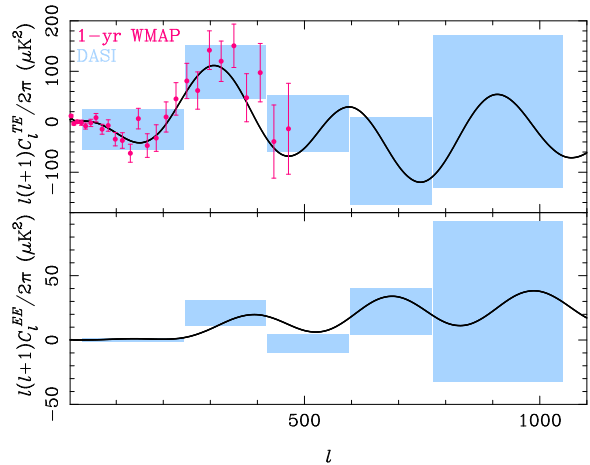}
\end{center}
\caption{Current measurements (as of February 2004) of $C_l^{TE}$ (top) and
$C_l^E$ (bottom). The points with 1-$\sigma$ errors are from the first
one-year data release from WMAP~\cite{adc:hinshaw03}.
The error boxes are the flat band-power results from DASI~\cite{adc:kovac02}
centred on the maximum-likelihood
band power and spanning the 68-per cent interval. The solid lines are the
predicted power from the best-fit model to all the WMAP data.
}
\label{adc:fig14} 
\end{figure}

The first detection of polarization of the CMB was announced
in September 2002~\cite{adc:kovac02}. The measurements were made with DASI,
a compact interferometric array operating at 30\,GHz, deployed at the
South Pole. The DASI team
constrained the amplitude of the $E$ and $B$-mode spectra
with assumed spectral shapes derived from a concordant
$\Lambda$CDM model. They obtained
a $\sim 5$-$\sigma$ detection of a non-zero amplitude for $E$ with a central
value perfectly consistent with that expected from the amplitude of
the temperature anisotropies. DASI also detected the temperature--polarization
cross-correlation at 95-per cent significance, but
no evidence for $B$-mode polarization was found.
The DASI results of a maximum-likelihood band-power estimation of the $E$ and
$TE$ power spectra are given in Fig.~\ref{adc:fig14}.

Measurements of $C_l^{TE}$ were also provided in the first-year data
release from WMAP, although polarization data itself was not
released. These results are also shown in Fig.~\ref{adc:fig14}. The existence
of a cross-correlation between temperature and polarization on degree
angular scales provides evidence for the existence of super-horizon
fluctuations on the last scattering surface at recombination. This is more
direct evidence for such fluctuations than from the large-scale temperature
anisotropies alone, since the latter could have been generated gravitationally
all along the line of sight. The sign of the cross-correlation and the
phase of its acoustic peaks relative to those in the temperature-anisotropy
spectrum is further strong evidence for adiabatic fluctuations. The one
surprise in the WMAP measurement of $C_l^{TE}$ is the behaviour on large
scales. A significant excess correlation over that expected
if polarization were only generated at recombination is present on large
scales ($l < 20$). The implication is that reionization occurred early,
$11 < z_\mathrm{re} < 30$, giving a significant optical depth for
re-scattering: $\tau_\mathrm{re} = 0.17 \pm 0.04$ at 68-per cent confidence.
As mentioned in Sect.~\ref{adc:subsec:other}, reionization at this epoch is
earlier than that expected from observations of quasar absorption spectra and
suggests a complex ionization history.

\subsection{Implications of Recent Results for Inflation}

The generic predictions from simple inflation models are that: (i) the
universe should be (very nearly) spatially flat; (ii) there should
be a nearly scale-invariant spectrum of Gaussian, adiabatic density
perturbations giving apparently-super-horizon fluctuations on the
last scattering surface; and (iii) there should be a stochastic background
of gravitational waves with a nearly scale-invariant (but necessarily not
blue) spectrum. The amplitude of the latter is a direct measure of the
Hubble rate during inflation, and hence, in slow-roll models, the energy
scale of inflation.

As discussed in Sect.~\ref{adc:subsec:curvature},
the measured positions of the acoustic peaks constrains the universe to
be close to flat. The constraint improves further with the inclusion of
other cosmological data. 
There is no evidence for isocurvature modes
in the CMB, although the current constraints are rather weak if general,
correlated modes are allowed in the analysis~\cite{adc:bucher04}.
Several of the
cosmological parameters for the isocurvature models most favoured by CMB data
are violently at odds with other probes, most notably the baryon density which
is pushed well above the value inferred from the abundances of the
light-elements. There is also no evidence for primordial non-Gaussianity in
the CMB (see e.g.~\cite{adc:komatsu03})\footnote{The WMAP data does appear
to harbour some statistically-significant departures from rotational
invariance~\cite{adc:olcosta03,adc:vielva03,adc:copi03,adc:eriksen04,adc:hansen04}. The origin of these effects, i.e.\ primordial or systematic
due to instrument effects or imperfect foreground subtraction, is as
yet unclear.}.

Within flat $\Lambda$CDM models with a power-law spectrum of curvature
fluctuations, the spectral index is constrained by the CMB to be close to scale
invariant~\cite{adc:spergel03},
although the inclusion of the latest data from small-scale experiments,
such as CBI~\cite{adc:readhead04} and VSA~\cite{adc:rebolo04},
tends to pull the best fit from
WMAP towards redder power-law spectra: e.g.\ $n_\mathrm{s} =
0.97^{+0.06}_{-0.03}$ at 68-per cent confidence combining WMAP and
VSA~\cite{adc:rebolo04}.
Slow-roll inflation predicts that the fluctuation spectrum
should be close to a power law, with a run in the spectral index that is
second order in slow roll: $\D n_\mathrm{s} / \D \ln k \sim
(n_\mathrm{s}-1)^2$.
The WMAP team reported weak evidence for a running spectral index by including
small-scale data from galaxy redshift surveys and the Lyman-$\alpha$ forest,
but modelling uncertainties in the latter have led many to question the
reliability of this result (e.g.~\cite{adc:seljak03}).
New data from CBI and VSA
now provide independent evidence for running in flat $\Lambda$CDM models
at the 2-$\sigma$ level from the CMB alone. This reflects the tension between
the spectral index favoured by the low-$l$ CMB data (which is anomalously
low for $l<10$, favouring bluer spectra) and the high-$l$ data from the
interferometers. The evidence for running is weakened considerably
with the inclusion of external priors from large-scale structure data. The
best-fit values for the run in $n_\mathrm{s}$ obtained with the CMB alone
are uncomfortably large for slow-roll inflation models, and give low
power on small scales that is difficult to reconcile with the early
reionization implied by the WMAP polarization data.
However, a recent analysis~\cite{adc:slosar04} argues that the
evidence for running depends crucially on the techniques employed to
estimate the low-$l$ power from WMAP data, and that the running is strongly
suppressed if exact likelihood techniques are adopted. A definitive answer
on whether departures from power-law spectra are significant must probably
await further data on both large and small scales.

The final prediction of slow-roll inflation -- the generation of nearly
scale-invariant background of gravitational waves -- is yet to be verified.
The current limits on the tensor-to-scalar ratio are only
weak: \cite{adc:rebolo04} report $r < 0.68$ at 95-per cent confidence from all
CMB data in general, non-flat, adiabatic $\Lambda$CDM models. Despite this,
observations are beginning to place interesting constraints on specific
models of inflation in the $r$--$n_\mathrm{s}$
plane~\cite{adc:liddle03,adc:tegmark03}. Already,
large-field models with power-law potentials steeper than $V \propto \phi^6$
are ruled out due to their red scalar spectra and comparatively
large tensor-to-scalar ratio. Future programmes targeting $B$-mode
polarization may ultimately be able to detect gravitational waves down
to an inflationary energy scale of a $\mathrm{few}
\times 10^{15}\, \mathrm{GeV}$.
Such observations will sharpen constraints in the $r$--$n_\mathrm{s}$
plane considerably, and should allow fine selection amongst the many
proposed models of inflation.

\subsection{Detection of Late-Time Integrated Sachs-Wolfe Effect}

The late-time ISW effect arises from the decay of the
gravitational potentials once the universe becomes
dark-energy dominated, and so should produce large-angle (positive)
correlations between the CMB temperature anisotropies and other tracers of
the potential
in the local universe. With the advent of the WMAP data, a number of
groups have reported the detection of such a correlation.
In~\cite{adc:boughn04},
WMAP data was cross-correlated with data on the hard X-ray background
(which is dominated by emission from active
galaxies) from the HEAO-1 satellite, and the number density of radio sources
from the NVSS catalogue.
In each case a positive correlation was detected at significance
$3\sigma$ and $2.5\sigma$ respectively. The correlation with NVSS has also
been carried out independently by the WMAP team~\cite{adc:nolta03},
who also note
that the observed positive correlation can be used to rule out the
closed, $\Lambda=0$ model model that is a good fit to the CMB data in
isolation (see Fig.~\ref{adc:fig11}). Several groups have now also detected
the cross-correlation on large scales between the CMB and optical galaxy
surveys, e.g.~\cite{adc:fosalba03}.

\section{Conclusion}
\label{adc:sec:summary}

The linear anisotropies of the cosmic microwave background have been studied
theoretically for over three decades. The physics, which is now well
understood, employs linearised radiative transfer, general relativity,
and hydrodynamics to describe the propagation of
CMB photons and the evolution of the fluid constituents
in a perturbed Friedmann-Robertson-Walker universe.
A number of bold predictions have emerged from this theoretical activity,
most notably the existence of acoustic peaks in the anisotropy power spectrum
due to oscillations in the photon--baryon plasma prior to recombination.
Observers have risen to the challenge of verifying these predictions,
and their detection is proceeding at a staggering rate. The large-scale
Sachs--Wolfe effect, acoustic peak structure, damping tail, late-time
integrated Sachs--Wolfe effect, polarization and reionization signature
have all been detected, and the first three have been measured in considerable
detail. Already, the size and scale of these effects is allowing
cosmological models to be constrained with unprecedented precision. The
results are beautifully consistent with almost-scale-invariant adiabatic
initial conditions evolving passively in a spatially flat, $\Lambda$CDM
universe.

Much work still remains to be done to exploit fully the information
contained in the CMB anisotropies. The Planck satellite, due for launch
in 2007, should provide definitive mapping of the linear CMB anisotropies,
and a cosmic-variance limited measurement of the power spectrum up
to multipoles $l\sim 2000$. This dataset will be invaluable in assessing
many of the issues hinted at in the first-year release of WMAP data,
such as the apparent lack of power on large scales and possible
violations of rotational (statistical) invariance. Prior to Planck, a number
of ground-based programmes should shed further light on the issue of
whether departures from a power-law primordial spectrum are required
on cosmological scales, and the implications of this for slow-roll inflation.
In addition, these small-scale observations will
start to explore the rich science of secondary anisotropies, due to e.g.\
scattering in hot clusters~\cite{adc:sz72} or
bulk flows modulated by variations in the electron density in the reionized
universe~\cite{adc:sz80,adc:ostriker86}, and
the weak lensing effect of large-scale structure~\cite{adc:blanchard87}.

Detections of CMB polarization are in their infancy, but we can expect rapid
progress on this front too. Accurate measurements of the power spectra of
$E$-mode polarization, and its correlation with the temperature anisotropies,
can be expected from a number of ground and balloon-borne experiments,
as well as from Planck. The ultimate goal for CMB polarimetry is to detect
the $B$-mode signal predicted from gravitational waves. This would
give a direct measure of the energy scale of inflation, and, when combined
with measurements of the spectrum density perturbations, place
tight constraints on the dynamics of inflation.
Plans are already being made for a new generation of polarimeters with
the large numbers of detectors and exquisite control of instrument
systematics needed to detect the gravity-wave signal if the energy scale
of inflation is around $10^{16}\,\mathrm{GeV}$. Ultimately, confusion
due to imperfect subtraction of astrophysical foregrounds and the effects
of weak lensing on the polarization limit will limit the energy scales
that we can probe with CMB polarization; see~\cite{adc:hirata03} and
references therein.

\section*{Acknowledgments}

AC acknowledges a Royal Society University Research Fellowship.

\end{document}